# Permutation Group Symmetry and Correlations: Constructing Trial Wavefunctions for Quantum Liquid States*

John J. Quinn
University of Tennessee, Knoxville

Incompressible quantum liquid states occur when the single particle angular momentum $\ell$ and the number N of electrons in the partially filled Landau level satisfy the equation $2\ell=\nu^{-1}N - C_\nu$. Here $\nu$ is the filling factor, and $C_\nu$ is a finite size shift. The value of $C_\nu$ can be obtained from a simple heuristic argument as well as from numerical diagonalization studies. A trial wave function $\Psi$ can always be written as the product of $F\{z_{ij}\}$ and $G\{z_{ij}\}$, where $F = \Pi_{i<j}z_{ij}$ is an antisymmetric Fermion factor due to the Pauli exclusion principle, and G is a symmetric correlation function caused by interactions. For the simple Laughlin correlated states at $\nu^{-1}$ equal to an odd integer 2m+1, G is equal to F raised to the 2m$^{th}$ power. Correlation functions for the "paired state" (at $2+ \nu = 5/2$) of the half-filled first excited Landau level, and for Jain states at $\nu =n (1+2n)^{-1}$ are considered. They involve partitioning the N particle system into subsets $g_1$, $g_2$, …, and introducing correlation factors $(z_{ij})$ between particles within subsets, and between particles belonging to different subsets in such a way that the highest power of any coordinate $z_i$ is less than or equal to $2\ell=\nu^{-1}N - C_\nu$. For the Jain state at $\nu=3/7$, $g_1$, $g_2$, and $g_3$ contain $N/3 - 2$, $N/3$, and $N/3 + 2$ particles respectively. Quasiparticle states and their correlations can be treated in a similar way if $C_\nu$ is increased or decreased by unity.

*Preliminary version of this work was presented as a contributed paper (session N23, paper 8) of the 2013 March meeting of the APS.

The fractional quantum Hall effect is the paradigm of a strongly interacting many body system. At very large values of the applied magnetic field $B_0$ there is only one relevant energy scale, the Coulomb scale $V_c = e^2/\varepsilon_0|\vec{r}_i - \vec{r}_j|$ where $\varepsilon_0$ is the background dielectric constant. This is due to the degeneracy of the single particle states within the partially occupied Landau level. Laughlin[1] correctly surmised that the avoidance of the most strongly repulsive pair states (referred to as Laughlin correlations between electrons) would result in an incompressible liquid state. Laughlin's celebrated wavefunction for the $\nu = (2m+1)^{-1}$ state is $\Psi(1,2 \ldots, N) = \Pi_{i<j} z_{ij}^{2m+1} \exp(\Sigma_k \mid z_k|^2 / 4\lambda^2)$, where $z_i$ is the complex coordinate of the $i^{th}$ particle, and $z_{ij} = z_i$-$z_j$. The exponential factor is ubiquitous and will sometimes be omitted altogether for brevity. The wavefunction is antisymmetric because it changes sign under transposition of any pair <ij>. In addition, it prohibits any pair <ij> from being in a pair state with pair angular momentum $L_2$ larger than $2\ell - (2m+1)$ (or any relative pair angular momentum $R_2 \equiv 2\ell - L_2$ less than 2m+1). Because the pair interaction energy $V_0(L_2)$ for electrons in LL0 increase monotonically as $L_2$ increases, the Laughlin wavefunction avoids the most repulsive pair states

Jain[2] introduced a very simple picture to describe the most robust incompressible quantum liquid (IQL) states of LL0. They occur at $\nu = n(1\pm2pn)^{-1}$ where n and p are positive integers. He suggested that each electron captured 2p flux quanta ($\varphi = \frac{hc}{e}$) of the magnetic field in a flux tube rigidly attached to it. The resulting object (electron plus 2p attached flux quanta) was called a composite Fermion (CF). In a mean field approximation (MFA), the electron charge is considered to be uniformly distributed and is cancelled by a neutralizing background of positive charge. The CF moves as a non-interacting particle in a magnetic field $B^* = B_0 - 2\varphi n_0$, where $n_0$ is the average number of electrons per unit area. Because $\nu^{-1}$ is the flux per electron associated with the field $B_0$, the effective flux per CF is given by $1/\nu^* = 1/\nu - 2p$. When $\nu^*$ is equal to an integer ($\nu^* = \pm n$) the CF Landau level has integral filling, and an incompressible daughter state results at electron filling factor $\nu = n(2pn\pm1)^{-1}$. This very simple picture accounts very well for the values of the electron filling factors $\nu$ at which the most robust IQL states in LL0 occur. In addition, it correctly predicts the lowest band of angular momentum multiplets for any value of $B_0$ by requiring it to contain the minimum number of QP excitations[3] consistent with the values of N and $2\ell$. Quasiholes have angular momentum $\ell_{QH} = \ell - (N-1)$ and quasielectrons $\ell_{QE} = \ell_{QH} + 1$. The lowest energy band of system containing $N_{QP}$ quasiparticles each with angular momentum $\ell_{QP}$, is obtained by simple addition of the angular momenta of $N_{QP}$ Fermion quasiparticle each with angular momentum $\ell_{QP}$.[3]

The success of the Laughlin correlated wavefunction and of Jain's CF picture rests on agreement with numerical diagalization studies of small systems (usually N less than 12 to 14)[4]. IQL states occur at the values of ν predicted by the Laughlin-Jain picture. For the Laughlin states, the wavefunction determined in exact numerical diagonalization studies can be compared with the Laughlin trial wavefunction. The overlap is found to be close to unity. No comparable trial wavefunctions for the Jain states at $\nu = 2/5, 3/7, \ldots, 2/3, 3/5, \ldots$, have been suggested which do not require extensive numerical calculations. The only other trial wavefunction suggested to describe an IQL state is the Moore-Read[5] wavefunction for the half-filled state of one spin orientation in the first excited Landau level (2+ν= 5/2). In the following sections we will discuss construction of trial wavefunctions for both Jain state at $\nu=n(1\pm2n)^{-1}$ and Moore-Read type "paired states."

Any trial wavefunction can be written as a product of an antisymmetric Fermion factor $F = \Pi_{i<j}z_{ij}$ resulting from the exclusion principle, and a symmetric correlation factor $G_L\{z_{ij}\}$ caused by the interactions. For the $\nu=(2m+1)^{-1}$ Laughlin wavefunction[1] $G_L = \Pi_{i<j}z_{ij}^{2m}$. For the Moore- Read wavefunction $G_{MR}$ is equal to F multiplied by the pfaffian $\mathrm{Pf}(z_{ij}^{-1}) = \hat{A}_N\{(z_{12}z_{34}\ldots z_{N-1,N})^{-1}\}$[5,6] . Here $\hat{A}_N$ stands for the operator which antisymmetrizes the product in curly brackets over all N particles. Since G is a product of two antisymmetric functions, it must be symmetric as required.

It is known that IQL states in LL0 occur at $\nu=n(2pn\pm1)^{-1}$ if the value of the shell angular momentum $\ell$ satisfies the relation $2\ell=\nu^{-1}N-C_\nu$. Here ν is the filling factor, N the number of electrons, and $C_\nu$ is the "finite size shift" for the Haldane spherical geometry. This relation is obtained in numerical diagonalization of N electron systems confined to a spherical surface of radius R with a magnetic monopole of strength 2Qφ at its center to produce a radial magnetic field $B_0 = \frac{2Q\varphi}{4\pi R^2}$. It can also be obtained without numerical calculation by a simple heuristic argument. For $\nu = n/q < 1/2$, the $2\ell+1$ single particle states are divided into $(N/n-1)$ unit cells, each containing q single particle states of which the first $n$ are filled, and the remainder empty. This accounts for $n(\frac{N}{n}$ -1) = N- $n$ of the electrons. The remaining n are added after the last unit cell. This results, for the case of $\nu = 2/5$, in the simple diagram of Fig. 1. It is clear that the total number of states $2\ell+1$ must be equal to $q(N/n-1)+n=(q/n)N-(q+1-n)$ giving[7] $C_\nu$=q+1-n.

It must be emphasized that the parent state depicted in Fig.1 is a non-interacting state, and is not an eigenstate of the interacting system. However, when interactions are turned on, many states with different occupancies of the same single particle levels will be generated. One linear combination of this myriad of states, all of which have $M_z$, the z component of the total angular momentum, equal to zero, will be the L = 0 IQL ground state. We propose to construct electronic trial wavefunctions by selecting symmetric correlation functions for which the maximum allowed value of any power of $z_i$ is $2\ell=\nu^{-1}N-C_\nu$. For the Laughlin state $G=\Pi_{i<j}(z_{ij})^{2m}$ giving $C_\nu$ = 2m+1 in agreement with $C_\nu$ = q+1- $n$ for $\nu$ = $n$ /q = $m^{-1}$. For the MR state, G = F · Pf($1/z_{ij}$) contains $\Pi_{i<j}z_{ij}$ from F and a large number of terms containing one factor $z_{lk}^{-1}$ for different values of k giving $z_l^{-1}$ for each of terms. Thus G contains $z_l^{(N-1)}z_l^{-1}$ and F contains $z_l^{(N-1)}$ giving $2\ell=2N-3$ as the maximum value of any $z_i$ in Ψ=FG. It is worth noting that simply choosing G by partitioning N into two subsets $g_A$ = (1,3,5,…,N-1) and $g_B$ = (2,4,6,…,N) allows one to write a symmetric quadratic correlation function $G_Q=\hat{S}_N$ $\{\Pi_{i<j\varepsilon g_A}(z_{ij})^2\Pi_{k<l\varepsilon g_B}(z_{k\ell})^2\}$, where $\hat{S}_N$ symmetrizes the product in curly brackets over all N particles (or over all partitions of N into $g_A$, $g_B$ each containing N/2 particles). This correlation function gives the same $2\ell=2N-3$ as Moore-Read, but it is different.

For Jain states at $\nu=n/q<\frac{1}{2}$ , we partition the N particle system into subsets $g_1,g_2,\ldots,g_n$, each containing two particles more than the preceding one. For example, for n=3, $g_1,g_2,g_3$ contain $(N/3)$ - 2, $N/3$, and $(N/3)$ + 2 respectively. Then we construct a symmetric correlation function such that the largest value of any power of $z_i$ is less than or equal to $2\ell=\nu^{-1}N-C_\nu$. There can be different correlations (powers of $z_{ij}$) between a pair of particles in one subset and a pair in a different subset, and still different correlations between particles belonging to different subsets.

For $\nu = {}^{2}/_{5}$, as an example, we partition N into $g_1$ and $g_2$ containing $\frac{N}{2}-1$ and $\frac{N}{2}+1$ particles respectively. For an N=6 particle system one partition is $g_1$= (1,2) and $g_2$=(3,4,5,6). We choose a symmetric correlation function:

$$G_{2/5} = [F(z_{ij})]^2\, \hat{S}_6[f(p_{12})]$$

Again $\hat{S}$ symmetrizes $f(p_{12})$ over all six particles, and $f(p_{12})$ can be written:

$$f(p_{12})=T_{12}U_{12}$$

The functions $T_{12}$ and $U_{12}$ are defined by:

$$T_{12}= \Pi_{3\le p\le 6}(z_{1k}z_{2k})^{-1}$$
$$U_{12}=\hat{S}_4\{(z_{34}z_{45}z_{56}z_{63})^{-2}\}$$

The $\hat{S}_4$ appearing in $U_{12}$ symmetrizes the product in curly brackets over the four particles (3,4,5,6). It contains three terms involving only $z_3$ to $z_6$, which can be depicted as shown in Fig. 2. In this figure, lines connecting $z_i$ and $z_j$ stand for a factor $z_{ij}^{-2}$. There are $N!/(\frac{N}{2}-1)!\,(\frac{N}{2}+1)!$ partitions of the N particle system (15 for the N =6 particle system). The symmetric correlation function can be written:

$$G_{2/5} = [F(z_{ij})]^2\Sigma_{k<\ell}T_{k\ell}U_{k\ell}\ .$$

For the partition in which particles 1and 2 belong to $g_1$, the three (out of 45) diagrams contributing to the electronic wavefunction are displayed in Fig. 3. $\Psi_{2/5}$ for the six particle system can be expressed as:

$$\Psi_{2/5}=[F(z_{ij})]^3\hat{S}_6\{\Pi_{3\le j\le 6}(z_{1j}z_{2j})z_{12}^{2}\ [z_{35}^{2}z_{46}^{2}+z_{36}^{2}z_{45}^{2}+z_{34}^{2}z_{56}^{2}]\}$$

It is certainly not clear that the choice of $G_\nu\{z_{ij}\}$is unique. For the $2+\nu=5/2$ state (half-filled state of one spin orientation in LL1) we found two possible correlation functions, $G_{MR}\{z_{ij}\}$ and the quadratic $G_Q\{z_{ij}\}$. Numerical diagonalization studies will help to decide which of the possible correlation functions is most appropriate.

Not only IQL states, but states containing one or more quasiparticles can be treated in the same way. As one example we consider a single QE in a $\nu$=1/3 IQL state. We take $2\ell$ =3(N-1)-1 = 3N-4. In Jain's CF picture the effective CF angular momentum is given by $2\ell^*=2\ell-2(N-1)=N-2$. The lowest CF level can accommodate $2\ell^*+1$ = (N-1) composite Fermions, leaving one QE in the shell with $\ell_{QE}=\ell^*+1=(\frac{N}{2})$. We partition the N electron system into $g_1$ containing one electron and $g_2$ containing the remaining N-1. We propose an electronic wavefunction $\Psi_{QE}=F\{z_{ij}\}G\{z_{ij}\}$with :

$$G\{z_{ij}\}=F^2\hat{S}_N(P_k).$$

Here $P_k=(z_{k,1}z_{k,2}\ldots z_{k,k-1}z_{k,k+1}\ z_{k,k+2}\ \ldots\ z_{k,N})^{-1}$. $P_k$ is a product over all $j\neq k$ taken in cyclic order. This gives $\hat{S}_N\{P_k\}=\Sigma_k P_k$. From the definition of $P_k$ it is clear that it is symmetric under permutation of k with any $j\neq k$. The wavefunction can be illustrated with the diagram in Fig. 4 for an N=6 particle system with the partition in which $z_1 \varepsilon g_1$ and $z_k$ for $k\neq 1$ belongs to $g_2$. It should be noted that this picture is in agreement with Jain's picture of a CF quasielectron, and somewhat different from Laughlin's.

Numerical diagonalization studies of small ($N\leq 12$) systems are currently being carried out for comparison of the eigenvalues and eigenfunctions with the trial functions proposed in this work.

ACKNOWLEDGEMENT:

The author would like to thank Professor Joseph Macek and Dr. Rachel Wooten for helpful discussions on the work reported here.

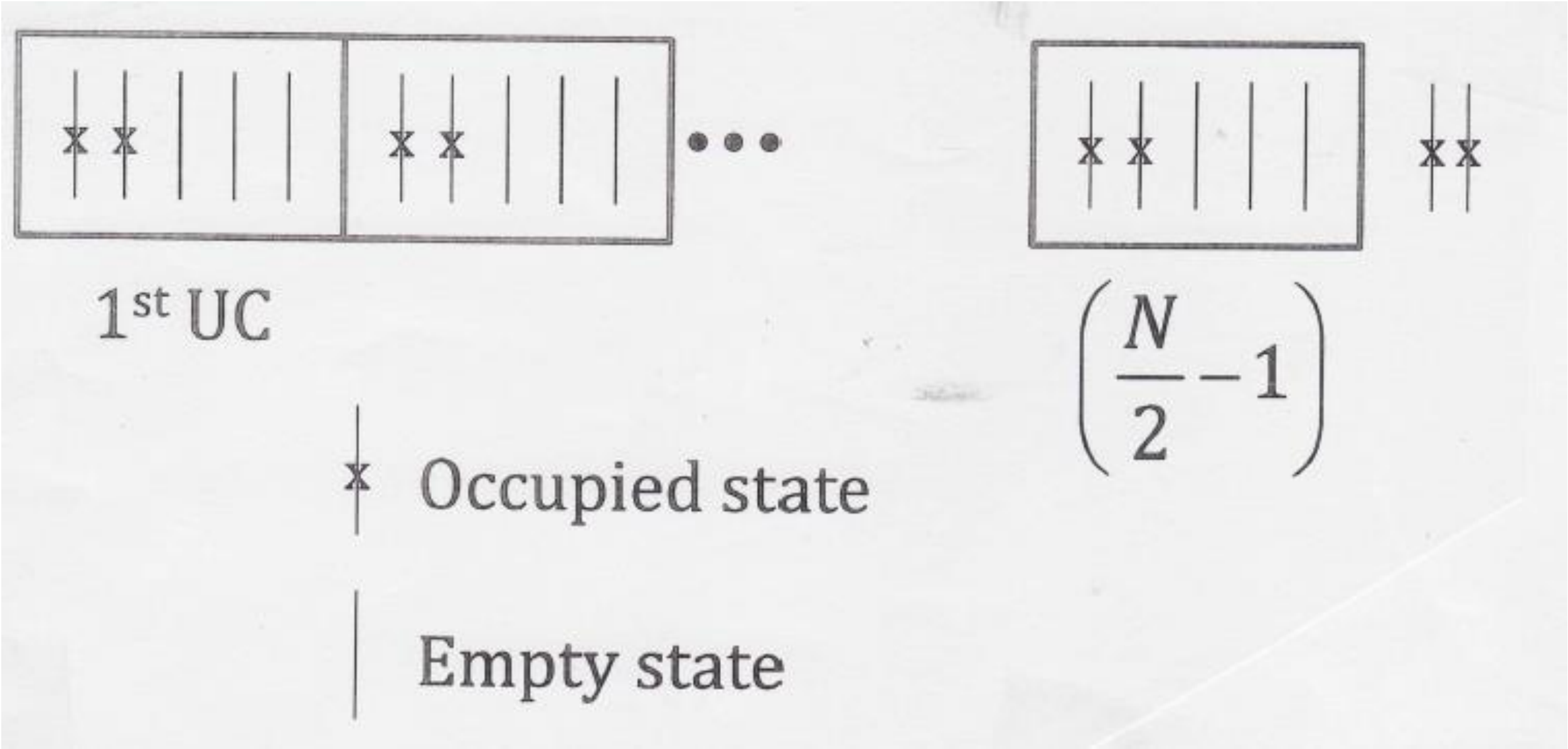

Figure 1. The 2ℓ+1 single particle states in a shell of angular momentum ℓ are represented by vertical lines having values which go from $\ell_z = -\ell$ to $\ell_z = +\ell$. Each unit cell (UC) contains five single particle states, the first two of which are occupied. There are ( $\frac{N}{2}$ -1) unit cells, and two filled single particle states after the last UC. This "parent state" has z-component of the total angular momentum L equal to zero. When interactions are turned on, it generates a myriad of states with other occupancies, all of which have total z-component $L_z$ of $\vec{L}$ equal to zero. One linear combination of this myriad of $L_z = 0$ states will be the IQL ground state.

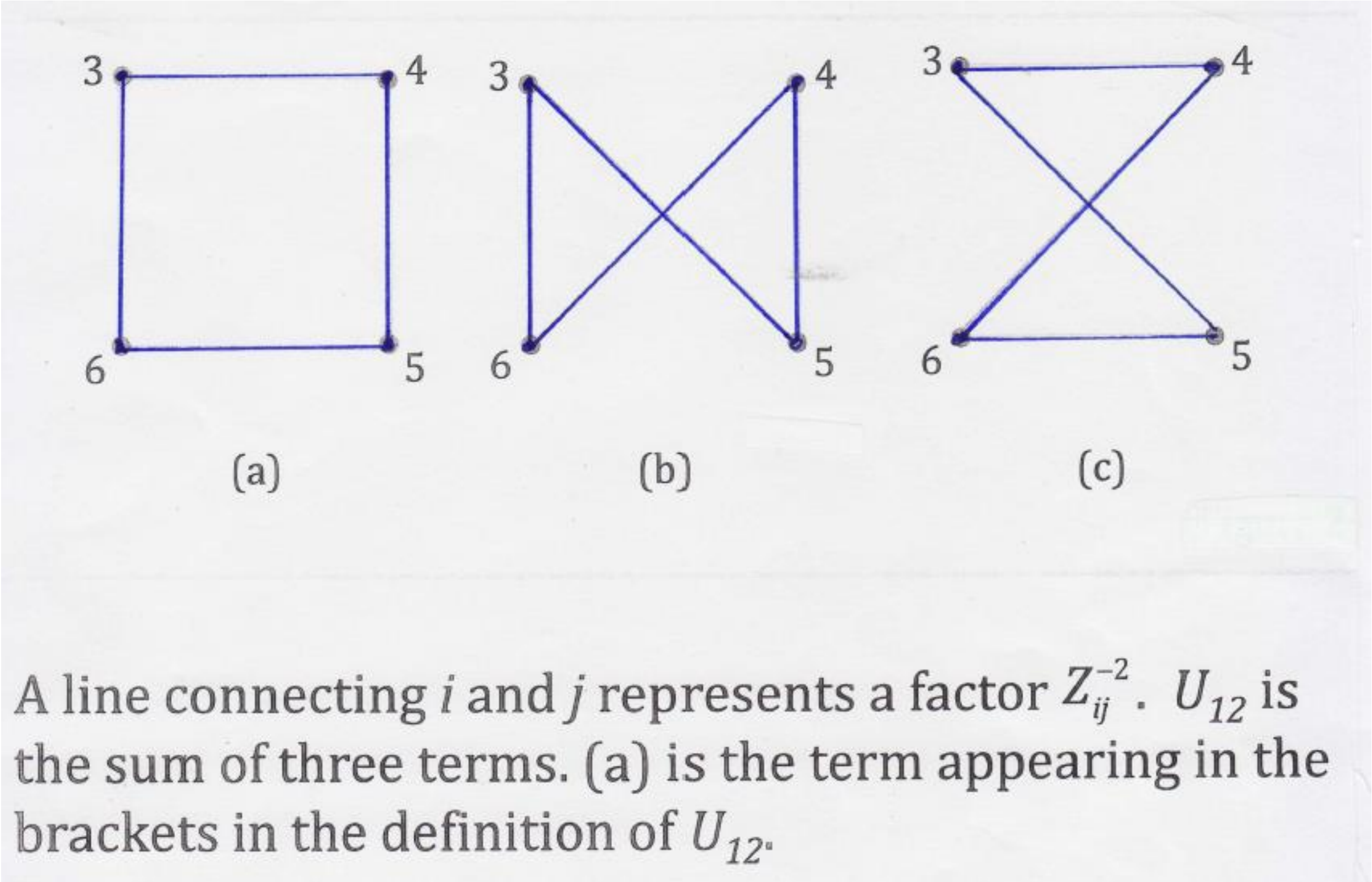

Figure 2. The three diagrams corresponding to $\hat{S}_4\{(z_{34}z_{45}\ z_{56}z_{63})^{-2}\}$ for the six particle system. Each line connecting i and j corresponds to a factor $z_{ij}^{-2}$.

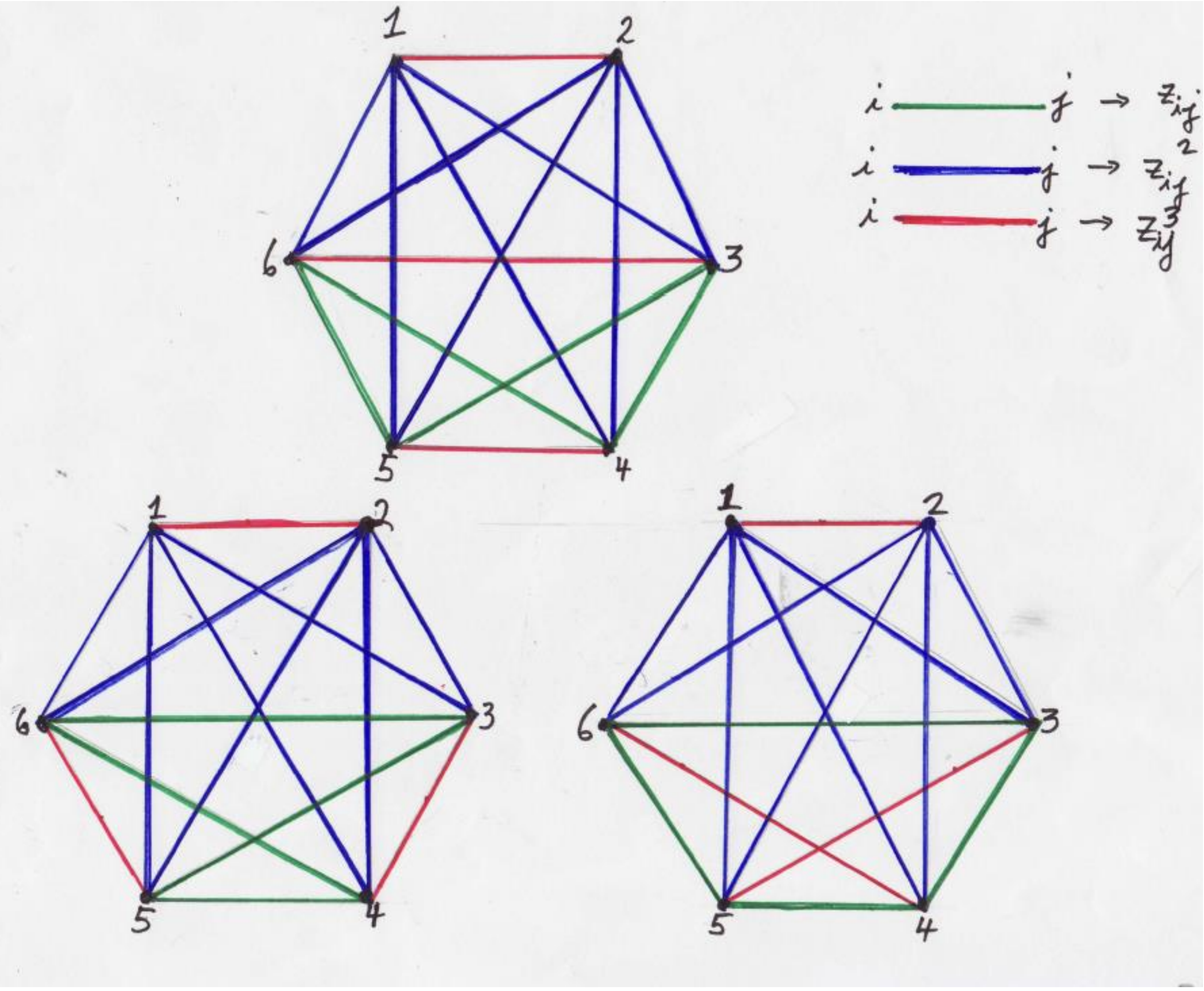

Figure 3. The three diagrams corresponding to the partition of $g_1 = \{1,2\}$ that contribute to the electronic trial wavefunction $\Psi_{2/5}$ . There are fifteen independent partitions of N = 6 into $g_A$ containing two particles and $g_B$ containing four. Each of these partitions gives rise to three diagrams. In the figure a red line between i and j indicates a factor $z_{ij}^3$, a blue line $z_{ij}^2$, and a green line $z_{ij}^1$.

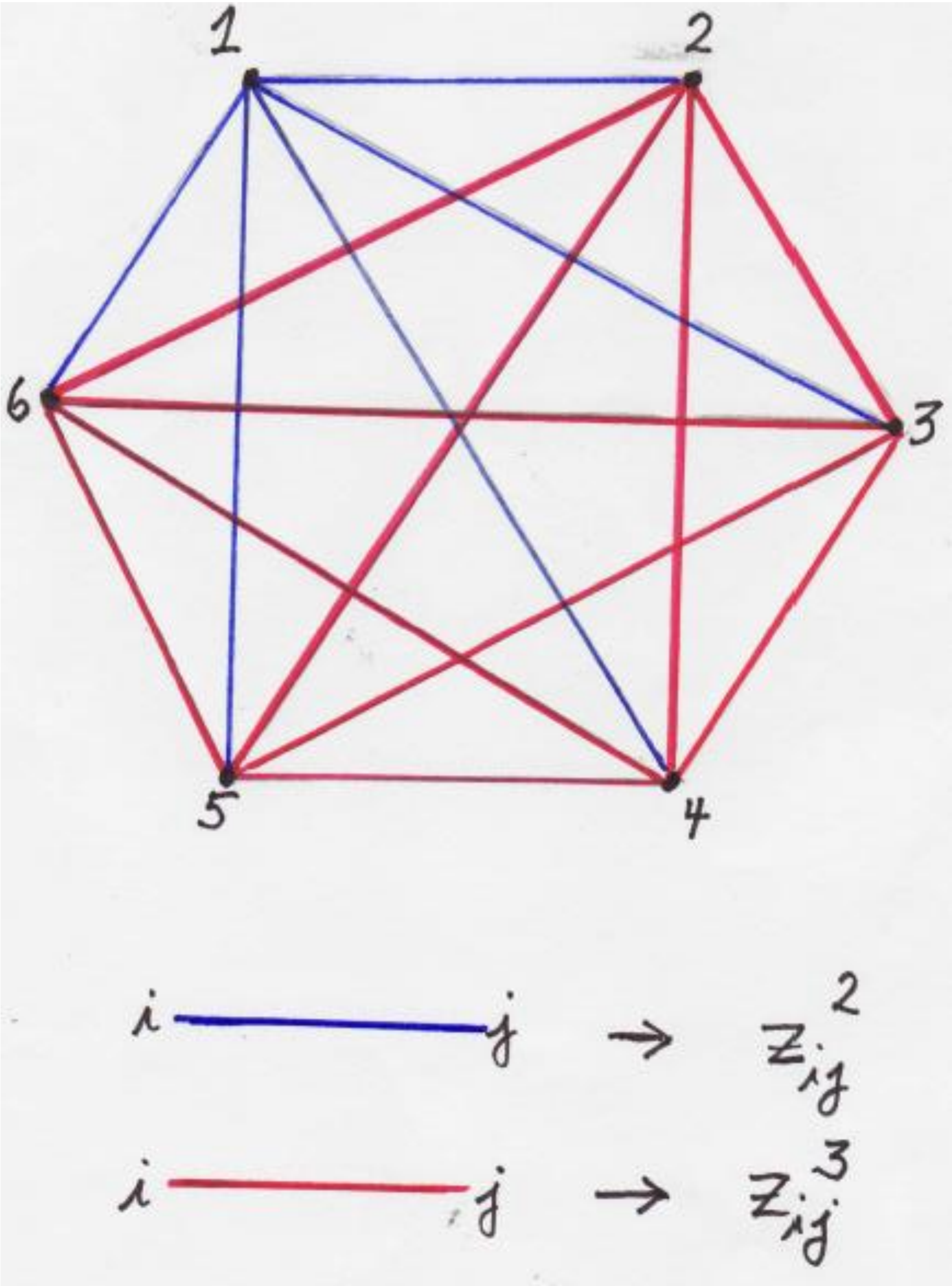

Figure 4. The diagram corresponding to the partition $g_1 = \{1\}$, $g_2 = \{2,3,4,5,6\}$of an $N = 6$ particle system in the single quasielectron wavefunction of a Laughlin $\nu=1/3$. Red lines between i and j indicate a factor of $z_{ij}^3$, blue lines a factor of $z_{ij}^2$. The trial function contains a sum over all partitions.